\documentclass[11pt, pra]{revtex4}
\usepackage{graphicx}% Include figure files
\usepackage{epsf}
\usepackage{dcolumn}% Align table columns on decimal point
\usepackage{bm}% bold math
\usepackage {amsmath, amssymb, amsfonts}

    \newcommand{\be}{\begin{equation}}
       \newcommand{\ee}{\end{equation}}
       \newcommand{\ba}{\begin{eqnarray}}
       \newcommand{\ea}{\end{eqnarray}}

\begin{document}

       \title{Investigating the Performance of an Adiabatic Quantum Optimization Processor}

 \author{Kamran Karimi, Neil G.~Dickson, Firas Hamze,  M.H.S.~Amin, Marshall Drew-Brook, Fabian A.~Chudak, Paul I.~Bunyk, William G.~Macready, and Geordie Rose}

\affiliation{D-Wave Systems Inc., 100-4401 Still Creek Drive, Burnaby, B.C., Canada, V5C 6G9}
\email{{kkarimi, ndickson, fhamze, mhsamin, marshall, fchudak, pbunyk, wgm, rose}@dwavesys.com}

 \begin{abstract}
Adiabatic quantum optimization offers a new method for solving hard optimization problems. 
In this paper we calculate median adiabatic times (in seconds) determined by the minimum gap during
the adiabatic quantum optimization for an NP-hard Ising spin glass instance class
with up to $128$ binary variables. Using parameters obtained from a realistic
superconducting adiabatic quantum processor, we extract the minimum gap and matrix
elements using high performance Quantum Monte Carlo simulations on a large-scale
Internet-based computing platform. We compare the median adiabatic times with the
median running times of two classical solvers and find that, for the considered
problem sizes, the adiabatic times for the simulated processor architecture are about
4 and 6 orders of magnitude shorter than the two classical solvers' times. This shows
that if the adiabatic time scale were to determine the computation time, adiabatic
quantum optimization would be significantly superior to those classical solvers for
median spin glass problems of at least up to 128 qubits. We also discuss important
additional constraints that affect the performance of a realistic system.
 \end{abstract}

 \maketitle

\section{Introduction}

NP-hard optimization problems arise frequently in scientific and commercial
applications \cite{garey, appref1}. Their formal intractability, combined with the
significant value that can be generated by finding better techniques for solving
them, has led to a thriving research community searching for more effective and
useful algorithms. While much progress has been made, for many applications the best
current algorithms are not nearly good enough. An example of a very difficult
optimization problem is the Eternity II puzzle \cite{EternityII}.

Adiabatic Quantum Optimization (AQO) algorithms \cite{Kadowaki,farhi, AQCrev, ATA081,
MRTtheory} may be effective tools to tackle hard optimization problems, making it
important to study their performance potential. AQO algorithms are physics-inspired
approaches that can be used as components of both exhaustive and heuristic solvers.
In these algorithms, a Hamiltonian changes over time in such a way that after the
computation is complete, the system has a non-zero probability of being in its lowest
energy state, also known as its ground state. The ground state encodes the global
optimum of the problem being solved. The excited states correspond to solutions that
are not globally optimal. The $M^{th}$ excited state corresponds to the $(M+1)^{th}$
best solution.

AQO algorithms are not well understood. In particular it is very difficult to predict how their running time scales with problem size for any particular instance class of interest. Most of the work that has been done on this question focuses on the issue of whether, for asymptotically large problems, these algorithms require running times that scale polynomially or exponentially with problem size for some specific choice of solver parameters and instance class. Here we are not going to directly address these difficult issues, and approach the analysis of an AQO algorithm from a much simpler perspective.

Previous studies of the performance of adiabatic algorithms \cite{farhi, Pyoung1,
pYoung2} were merely concerned with the scaling and not the actual value of the time
in e.g., seconds for a realistic system. Moreover, in all those studies it is
assumed, for example, that the temperature of a real system can always be kept much
lower than the minimum energy gap between the ground state and first excited state,
and that the adiabatic time is the only relevant timescale. Here, we present results
suggesting that under the same assumptions, a superconducting hardware implementation
\cite{DW6} of an AQO algorithm could offer a speedup of several orders of magnitude
over state of the art conventional approaches. However, we show that for a real
processor with reasonable energy scales, the minimum energy gap is much smaller than
feasible temperatures, even for fairly small problems, and thus, the simulations
cannot accurately predict runtime scaling of such a processor.  We also find that the
adiabatic time is currently negligible compared to other necessary processes, such as
readout of the results and thermalization (allowing the processor to cool down after
a readout has been performed).

The rest of this paper is organized as follows. Section 2 presents relevant details of the simulated processor. Section 3 introduces the problems we solve and provides information about the techniques needed to perform our large-scale simulations. Section 4 presents the classical solvers used in this study and compares their performance with that of our simulated AQO processor. Section 5 concludes the paper.

\section{The simulated processor}

In order to calculate adiabatic times for a quantum computation, in units of seconds,
it is necessary to target a specific AQO processor architecture. In more
theory-oriented publications, no limitation is placed on the connectivity of the
architecture, but arbitrary connectivity becomes impractical to fabricate as the
number of qubits increases. Here, we model a processor architecture with finite
connectivity that has been fabricated.  All of the simulation parameters come from
measurements done on the processor in \cite{DW6}, so there were no free parameters.
This processor is a programmable superconducting integrated circuit
\cite{SCALABLE-CONTROL} designed to perform AQO with up to 128 pairwise-coupled
\cite{DW1} superconducting flux qubits \cite{DW6}. To our knowledge, there are no
other AQO processor architectures of comparable size under development.

For analyzing an AQO process of a closed system at zero temperature, the key details
are captured in the hardware's Hamiltonian. As with the aforementioned studies, in
order to perform the analysis, we make the assumption that the temperature of the
processor is much smaller than the minimum energy gap, effectively zero, and that
there is no disturbance from any outside environment.  This approximation is clearly
not exact, but is necessary, because of the infeasibility of simulating a non-trivial
open system.  The Hamiltonian we consider here is of the form \be H(s)= A(s) \;  H_I
+ B(s) \; H_P \label{H} \ee $H_I$ and $H_P$ are dimensionless, and $s=t/t_f$ is a
normalized time $0 \leq s \leq 1$. The initial dimensionless Hamiltonian as
implemented by the hardware \cite{DW6} is \be H_I=-\sum_{i=1}^N \sigma^x_i \label{HI}
\ee where $\sigma^x_i$ is the Pauli X matrix for qubit $i$.

\begin{figure}
\begin{center}
\includegraphics[width=0.55\textwidth]{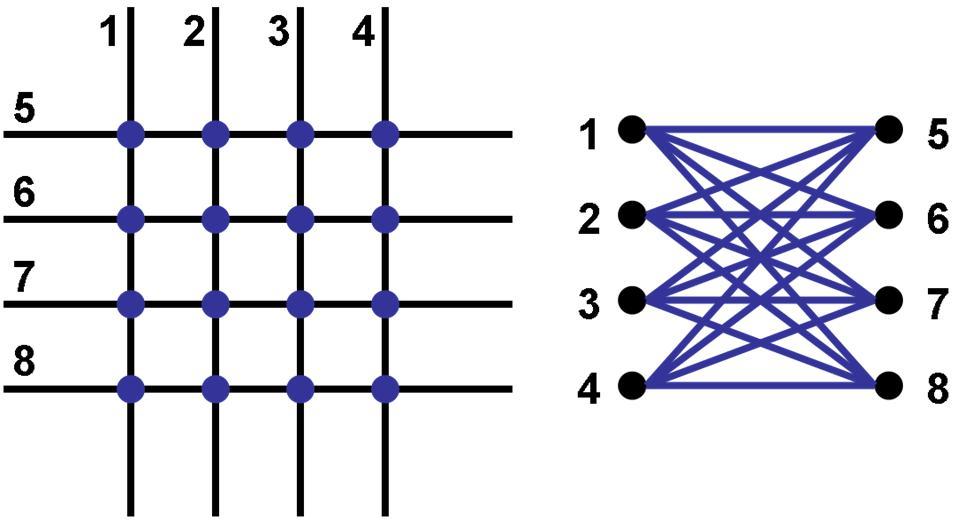}
\end{center}
\begin{center}
\includegraphics[width=0.55\textwidth]{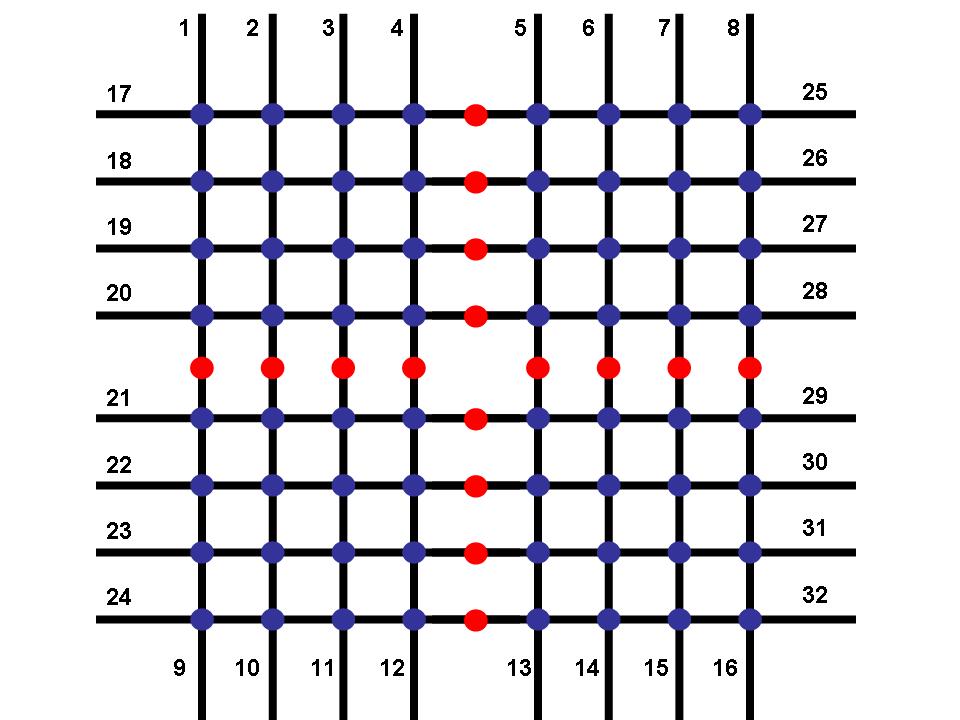}
\caption{(Colour online) Top: Two representations of the 8 vertex/variable/qubit unit cell graph (the complete bipartite graph on 4 vertices $K_{4,4}$). On the left, the variables are shown as black lines, with the allowed edge set $E$ denoted by the set of blue points (intra-cell couplers) where the lines intersect. Note that this is not a conventional graph representation. On the right, showing the conventional graph representation, the variables are the black dots, and the non-zero edges are shown as the blue connecting lines. Bottom: 4 tiled $K_{4,4}$ unit cells, giving a total of 32 variables. Here the red circles represent edges between unit cells (inter-cell couplers). Larger edge sets are constructed by further tiling.} \label{fig1}
\end{center}
\end{figure}

The final Hamiltonian, $H_P$, encodes the problem to be solved, and must be diagonal in the basis in which the qubits are read out.  Because of this, the final state of the system will be classical, and therefore, qubits can be read out individually without destroying the state. The specific form we use is
 \be
 H_P=\sum_{i=1}^N h_i \sigma^z_i + \sum_{i,j \in E} J_{ij} \sigma^z_i \sigma^z_j
 \label{HP}
 \ee
where $\sigma^z_i$ is the Pauli Z matrix for qubit $i$. The real-valued vector $h$
and matrix $J$ together define a {\em problem instance}, and are normalized such that
$-1 \leq \{h_i, J_{ij} \} \leq +1$. $E$ is the {\em allowed edge set}, dictating
which $J_{ij}$ can be non-zero (all other $J_{ij}$ are zero). In other words, $E$ is
the set of pairwise connections between qubits in our particular processor
\cite{DW6}. We restrict $E$ to correspond to tiled bipartite $K_{4,4}$ graphs, as
shown in Fig.~\ref{fig1}.

\begin{figure}
\begin{center}
\includegraphics[width=0.75\textwidth]{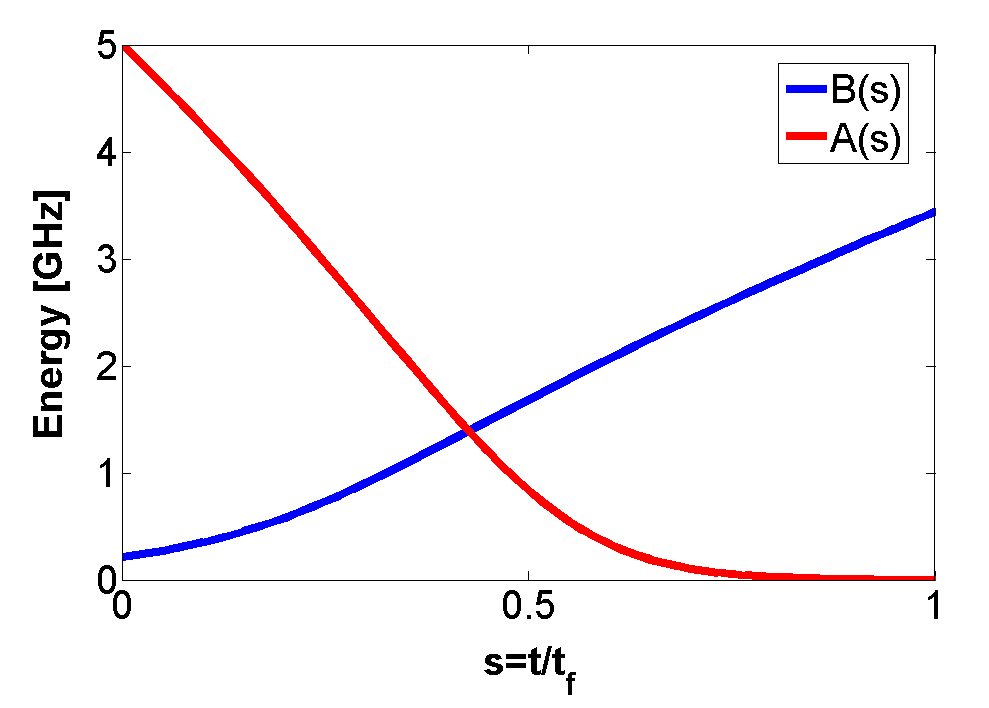}
\caption{(Colour online) Envelope functions $A(s)$ and $B(s)$ of the simulated processor. \label{fig2}}
\end{center}
\end{figure}

The functions $A(s)$ and $B(s)$ in Eq.~(\ref{H}) have units of energy. Here we use
the functional forms shown in Fig.~\ref{fig2}. These functions are determined from
the devices in the real processor we have simulated \cite{DW1, DW2, DW4, DW6}, i.e.
the processor uses a similar annealing schedule.  The energy scales of $A(s)$ and
$B(s)$ are limited by practical considerations for the actual superconducting devices
used in the modeled processor, such as junction sizes, critical currents, etc. At the
beginning of the computation (the leftmost side of Fig.~\ref{fig2} at $s = 0$), the
transverse term (\ref{HI}) in the Hamiltonian (\ref{H}) is dominant. As the
computation proceeds, the strength of the diagonal term  (\ref{HP}) grows while the
transverse term shrinks. Once the transverse term has been turned off completely (the
rightmost side of Fig.~\ref{fig2} at $s = 1$) the qubits are read out in the Z basis.

 Note that while these are not the envelope functions $A(s) \sim (1-s)$ and $B(s) \sim s$ used in other studies of AQO \cite{farhi, Pyoung1}, they do satisfy the underlying requirement of the AQO algorithms in that $A(0) \gg B(0)$ and $A(1) \ll B(1)$. We do not expect the results to change significantly if the linear envelope functions were used.

Changing $s$ from $0$ to $1$ within time $t_f$ at zero temperature results in finding
the global optimum with measurable probability as long as $t_f \gtrsim t_a$, where
\cite{childs, SQD}
 \be
 t_a = {{4 \left| <1|{{dH}\over{ds}}|0> \right|_{s=s^*}}\over{\pi g_m^2}},
 \label{adiabatic}
 \ee
which here we refer to as the adiabatic time scale. $g_m$ is the minimum energy gap
between the ground and first excited states of the Hamiltonian (\ref{H}) and occurs
at parameterized time $s^*$, which is unknown for any particular choice of $\{h,J\}$. Calculating $t_a$  for a general, 
arbitrarily large problem may be very challenging, if at all possible. In such cases different
values of $t_f$ may be tried in order to obtain an acceptable rate of success.

\section{Calculating the adiabatic times}

We wish to calculate the performance of the AQO algorithm embodied in Eq.~(\ref{H}) on a group of problem instances ranging in size from 8 to 128 variables. Table \ref{table1} describes the tiling of bipartite $K_{4,4}$ graphs used to structure problem instances of different sizes.

It is necessary to choose a scheme for generating problem instances where these
instances are all similar in some way. Here, we generate spin glass instances by
randomly assigning all $h_i$ and intra-cell couplers $J_{ij}$ to be either $+1/3$ or
$-1/3$ with equal probability. We set all inter-cell couplers to be $J_{ij}=-1$. We
also require each instance to have a unique global minimum, so instances found to
have multiple global minima were removed.
Solving for the ground state of Eq.~(\ref{H}) at $s=1$ using the $\{h,J\}$ generated
using this prescription is an NP-hard optimization problem \cite{baharona, istrail}.
For each problem size $N$, we randomly generated 100 problem instances. All
problem instances used in this work are available online \cite{onlineinstances}.

\setlength{\belowcaptionskip}{10pt}
\begin{table}
\begin{center}
\caption{The first three columns show the arrangement of unit cell graphs (shown in Fig. \ref{fig1}), and the resulting total number of variables in the examined problem instances. 8- and 32-variable instances are as shown in Fig.~\ref{fig1}. Input to classical solvers were problems of these sizes. For simulations, the average number of $s$ values for each problem size and the corresponding total number of variables in the simulated system are shown in the last two columns. The AQO simulations solved problems of these sizes. All simulations were performed with 256 Trotter slices.}\label{table1}

    \begin{tabular}{ | c |  c  | c | c | c | c |}
    \hline
    \# Rows & \# Columns & \# Variables  & \# $s$ Values  & Simulated Variables\\ \hline
    1 & 1 & 8 & 30 & 61,440 \\ \hline
    2 & 1 & 16 & 40 & 163,840 \\ \hline
    2 & 2 & 32 & 50 & 409,600 \\ \hline
    3 & 2 & 48 & 60 & 737,280 \\ \hline
    3 & 3 & 72 & 70 & 1,290,240 \\ \hline
    4 & 3 & 96 & 110 & 2,703,360 \\ \hline
    4 & 4 & 128 & 130 & 4,259,840 \\ \hline
 \end{tabular}
\end{center}

\end{table}

\setlength{\belowcaptionskip}{0pt}

Calculation of the adiabatic time can be done by exact diagonalization for small
problems (e.g. 8 or 16 variables), but not for larger problems, due to the
exponential growth of the Hamiltonian. We use a discrete imaginary time quantum Monte
Carlo (QMC) approach, as described in \cite{Pyoung1}, to compute the minimum gap and
matrix elements necessary for calculating the adiabatic time at all problem sizes
considered in this paper. We apply the Suzuki-Trotter decomposition \cite{Suzuki76}
to be able to use classical Monte Carlo techniques (e.g. Metropolis) to perform the
simulations. As described in \cite{Pyoung1}, doing so introduces an effective temperature 
to the resulting system, which does not negatively affect the simulation, so long as this effective 
temperature is well below the minimum gap.

To perform our adiabatic time calculations, it was necessary to simulate
the Hamiltonian (\ref{H}) at different values of \( s \) (e.g. about 130 different
values for the biggest problems we consider). Table \ref{table1} shows the average
number of $s$ values for problems of different size. For values of \( s \) close to
1, however, the resulting systems were such that single-site Monte Carlo algorithms
suffered from severe equilibration issues. A straightforward \emph{parallel
tempering} (PT) \cite{Hukushima96} solution is to supplement the single-site dynamics
with \emph{exchange} moves of spin configurations between systems at adjacent values
of \( s \). This differs slightly from the way PT is usually applied to classical
spin systems, where the simulations take place at different \emph{temperatures}; in
our context, the small and large \(s \) values play analogous roles to high and low
temperatures respectively.

For systems with first-order phase transitions \cite{pYoung2}, it is not sufficient to merely ensure that adjacent systems' parameters are sufficiently close
together to yield a reasonable (e.g. \( 25\%\) ) swapping probability. This problem was solved using an improved version of the \emph{feedback-optimized PT}  \cite{Katzgraber06} technique, as explained in \cite{Hamze1}. Without such a technique, obtaining reliable statistics for many of our systems would have been impossible.

In order to obtain the adiabatic times for AQO, each problem instance was simulated about 100 times, for a total of about 70,000 (7 problem sizes $\times$ 100 instances per size $\times$ 100 solutions per instance) simulations. For a typical 128-qubit problem, we needed to perform millions of Monte Carlo sweeps on about 4,259,840 (128 qubits $\times$ 256 Trotter slices $\times$ 130 $s$ values) variables, which required huge computing resources. We used multi-threading to effectively use multi-core CPUs and speed up the execution of the simulation program \cite {Karimi1}. To increase the performance of each thread, we vectorized the core of the Monte-Carlo method using Streaming SIMD Extensions 2 (SSE2) instructions. As a result of these and other optimizations, on an Intel Core i7-965 running at 3 GHz, the performance of our application increased about 50-fold compared to the single-threaded, unoptimized version.

 We used the Berkeley Open Infrastructure for Network Computing (BOINC) \cite{boinc} as a distributed computing platform. The resulting BOINC project, AQUA@home (Adiabatic QUantum Algorithms) \cite{aqua}, deployed our simulation application onto approximately 3,500 volunteer computers with a total of about 8,000 cores for around 3 months. Approximately $10^{20}$ floating point and integer operations were performed during the course of these computations. The simulation results, needed for computing Eq.~(\ref{adiabatic}), are available online for all $700$ instances studied \cite{online2}. We verified our simulations for 8- and 16-variable problems by comparing the simulation results with exact values obtained from diagonalization.

\begin{figure}
\begin{center}
\includegraphics[width=0.95\textwidth]{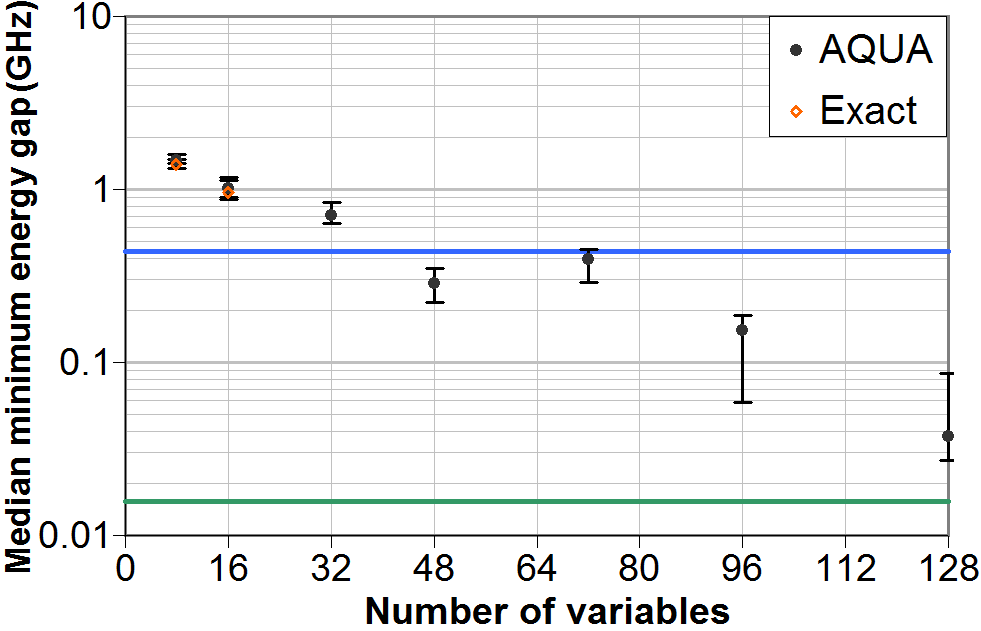}
\caption{(Colour online) Median minimum energy gaps as a function of problem size. Each data point is the median of the minimum energy gaps of 100 randomly generated instances. Error bars represent the $40\%$ and $60\%$ intervals around the median. The green line shows the simulation temperature of 0.75 mK (0.015625 GHz). The blue line shows the lowest feasible temperature of the real processor, 21 mK (0.44 GHz).} \label{fig3}
\end{center}
\end{figure}

 Similar to other AQO publications \cite{farhi, Pyoung1, pYoung2}, we use median values in our performance evaluations. The main reason is that the median of minimum energy gaps and running times are more useful than worst-case values, as they apply to typical problems. Notice that problems specifically designed to be hard (such as encryption) fall beyond our analysis.  Another important reason for using median values is that the Quantum Monte Carlo (QMC) simulation technique we used is not able to determine the numerical values of extremely small energy gaps, making it nearly impossible to investigate worst-case problems.

In Fig.~\ref{fig3} we plot the median minimum energy gaps as a function of problem
size. The simulation temperature of 0.75 mK (0.015625 GHz) is shown in green. We see
that even for the 128-variable problems, the median minimum energy gaps are larger
than this temperature. This implies that for the considered problems, the extraction of
ground state properties from the QMC simulations is reliable. The appropriateness of
using the median values is confirmed by considering the error bars on the gathered
data. We observe that running 100 instances provides a tight clustering of minimum
gap values and running times.

Fig.~\ref{fig3} appears to show that the median gap size for the 72-qubit instances is larger than that of the 48-qubit instances.  However, due to the nonzero error bars, this observation may not be correct.  Also, while is it very common for median gap sizes to decrease overall with respect to the problem size, there is no evidence to suggest that for these or similar problem instances, the median gap size should always decrease monotonically with system size. 

In the same figure, we plot (blue line) the lowest physical temperature, i.e., 21 mK,
that can be realistically achieved by the processor under normal conditions in a
dilution refrigerator. As can be seen, the median gap starts to get below the
temperature at around 48 variables. Even for the smallest sizes, the median gap is
only a factor of 2 to 3 larger than $T$. Therefore, the assumption of $T \ll g_m$,
commonly made in most calculations related to adiabatic quantum computation is not
realistic.

\section{Comparison with classical solvers' performance}
We compare our AQO implementation's adiabatic running time performance with that of two conventional solvers: IBM's CPLEX \cite{CPLEX}, and MadCat. CPLEX is a well-known commercial solver for discrete optimization problems, and using it provides a baseline for a general optimizer's performance. MadCat is a complete solver developed in-house, and designed to exploit the known structure of problem instances. MadCat is a particular case of a well-known general algorithm \cite{Kask} and relies on a \emph{tree decomposition}: a rooted tree whose nodes are subsets of the problem's variables, subject to certain conditions. Optimal assignments of variables in a given node can be determined, conditioned on the values of variables in the parent node. At the root node, unconditional optimal assignments are found and this information is propagated out to the leaves to produce a full optimal solution. The size of the largest tree node, minus 1, is called the \emph{treewidth} of the decomposition. The dynamic programming approach reduces MadCat's running time to be exponential in the treewidth, rather than exponential in the total number of problem variables. The $M \times N$ tiled graphs considered here have treewidth $\min(4M,4N)$, so the small instance sizes we consider here can be solved quickly. We also used MadCat to test the non-degeneracy of the problem instances.

CPLEX searches the space of solutions less efficiently than MadCat. To speed-up CPLEX's searches, we solved each instance using MadCat and obtained the optimal solution's energy. We then ran CPLEX with that energy as an argument to indicate that it should stop as soon as it finds a solution with that energy, i.e. as soon as it finds the optimal solution.

\begin{figure}
\begin{center}
\includegraphics[width=0.95\textwidth]{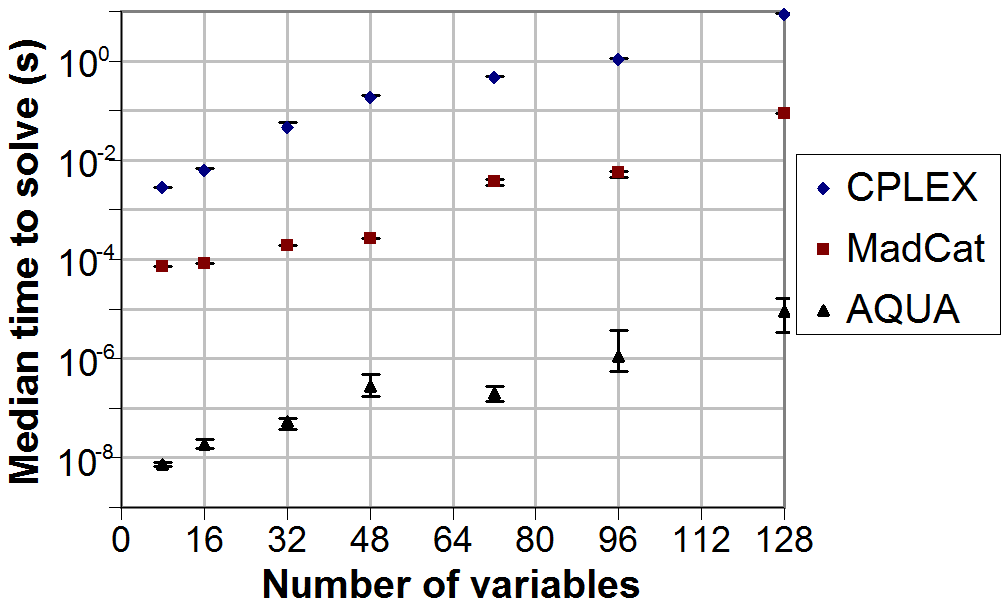}
\caption{(Colour online) Black: Median adiabatic times as a function of problem size. Each data point is the median of the adiabatic times of 100 instances.  Red (Blue): Median times to find the global optimum for MadCat (CPLEX). Error bars represent $40\%$ and $60\%$ intervals around the median.} \label{fig4}
\end{center}
\end{figure}

Fig.~\ref{fig4} shows the computed median adiabatic times, as well as the median running times required by both CPLEX and MadCat to find the global optimum of the same 700 instances. Both classical solvers were run on a system with 2 quad-core Xeon E5430 2.66 GHz processors running 64-bit Linux. CPLEX was set to use all 8 cores. Each instance was solved 10 times for both solvers, and of those, the minimum running time was retained to make sure that non-deterministic behaviour (for CPLEX), or transient system loads (for CPLEX and MadCat) did not negatively impact the measured running times.

Using different classical solvers, or different computer hardware for running them, would provide different results. However, since we already use reasonably high-performance software and hardware, we do not expect such changes to substantially affect the paper's conclusions.

It is important to emphasize that the predictions of running times for the AQO processor are not valid for problems with 48 or more variables running on the real processor at 21 mK, the lowest feasible temperature for the system. This is because the median minimum gap size drops below this operating temperature, as shown in Fig.~\ref{fig3}.  It is unknown, and non-trivial to determine, whether the true performance would be better or worse than the performance of a hypothetical processor at 0.75 mK \cite{SQD}.  If it is feasible to construct hardware with energy scales large enough that these median minimum gaps are increased to be larger than the operating temperature, such a processor would have significantly shorter adiabatic times than in Fig.~\ref{fig4}, because of Eq.~(\ref{adiabatic}).

\section{Concluding remarks}

This paper provides adiabatic running times for a specific AQO processor using
distributed high-performance simulations, and compares the results with those of two
classical solvers, i.e., CPLEX and MadCat. For the largest instances studied, MadCat
is roughly 2 orders of magnitude faster than CPLEX. This improvement is because
MadCat was designed to take advantage of the structure in the allowed edge set. Note
that the median running times for both solvers grow exponentially for larger problems
of this type. More striking is the computed median adiabatic time for the
superconducting adiabatic quantum processor, which is approximately 4 orders of
magnitude shorter than that of MadCat.  So, despite not knowing the performance
scaling of AQO for larger problems, we have shown that under the assumption that
adiabatic time correctly represents the time scale for computation, an AQO processor
could significantly outperform classical solvers.

Two important points have to be noted here. First, the above consideration ignores
the effect of the environment, especially the fact that the lowest feasible
temperature at which such a processor can be operated is larger than the median
minimum energy gaps of problems with 48 or more variables. Although recent
calculations suggest that a weak coupling to the environment does not significantly
affect the time required to reach the final ground state with a measurable
probability, even if the minimum gap is below the temperature
\cite{SQD,Childs01,Roland,Tiersch,Ashhab,Amin08,Wan,Amin09}, a fair comparison should
consider the performance of an open and not closed system. Unfortunately, such open
system simulations are not feasible beyond $\sim 20$ qubits \cite{SQD}.

The second point is that for the small problems we investigate here, the adiabatic
time does not dominate the running time of the real hardware.  A fair comparison
between the classical solvers and the AQO processor should include the time needed
for operations such as programming the chip, readout, and thermalization (to ensure
that the chip returns to its operational temperature before the next problem is
solved). In the real processor, readout and thermalization times completely dominate
the problem solving time for these small-scale problems. At the time of writing this
paper, serial readout takes roughly 36 $\mu s$ per qubit, and thermalization time is experimentally 
chosen to be $1000 \mu s$, as compared with median adiabatic time of $10 \mu$s for 128
variable problems. These numbers depend on the processors' design and fabrication
details, as well as the efficiency of cryogenic components used to cool the processor
to 21 mK, and are expected to change over time.

\section*{Acknowledgments}
The authors thank Edward Farhi, Helmut Katzgraber, and A. Peter Young for fruitful discussions. We would also like to enthusiastically thank the volunteer community supporting the AQUA@home distributed computing project, and the BOINC development team at UC Berkeley for their support.

 \end{document}